\documentclass[12pt]{iopart}
\usepackage{graphicx,epsfig}
\date{\today}

\begin{document}
\title[FES in general interacting systems]{Fractional exclusion statistics in general systems with interaction}
\author{Drago\c s-Victor Anghel}
\address{Department of Theoretical Physics, National Institute for
  Physics and Nuclear Engineering--''Horia Hulubei'', Str. Atomistilor
  no.407, P.O.BOX MG-6, Bucharest - Magurele, Romania}
\ead{dragos@theory.nipne.ro}
\begin{abstract}\noindent
I show that fractional exclusion statistics (FES) is manifested in general 
interacting systems and I calculate the exclusion statistics 
parameters. Most importantly, I show that the mutual exclusion statistics 
parameters--when the presence of particles in one Hilbert space influences 
the dimension of another Hilbert space--are 
proportional to the dimension of the Hilbert space on which 
they act. This result, although surprising and different from the 
usual way of understanding the FES, renders this statistics consistent 
and valid in the thermodynamic limit, in accordance with the 
conjucture introduced in J. Phys. A: Math. Theor. {\bf 40} F1013 (2007). 
\end{abstract}
\pacs{05.30.Ch,05.30.Pr}

\maketitle

\section{Introduction} \label{intro}

Fractional exclusion statistics (FES), introduced by Haldane in 
Ref. \cite{PhysRevLett.67.937.1991.Haldane} and with the thermodynamic 
properties calculated mainly by Isakov \cite{PhysRevLett.73.2150.1994.Isakov} 
and Wu \cite{PhysRevLett.73.922.1994.Wu}, has received very much attention 
since its discovery and has been applied to many models of interacting 
systems (see for example Refs. \cite{ProgrTheorPhys.5.544.1950.Tomonaga,JMathPhys.4.1154.1963.Luttinger,JMathPhys.6.304.1965.Mattis,PhysRevLett.81.489.1998.Carmelo,JMathPhys.10.2191.1969.Colagero,JMathPhys.12.247.1971.Sutherland,PhysRevA.4.2019.1971.Sutherland,PhysRevA.5.1372.1972.Sutherland,PhysRevLett.72.600.1994.Veigy,JPhysB33.3895.2000.Bhaduri,PhysRevB.60.6517.1999.Murthy,NuclPhysB470.291.1996.Hansson,IntJModPhysA12.1895.1997.Isakov,PhysRevLett.73.3331.1994.Murthy,PhysRevLett.74.3912.1995.Sen,PhysRevLett.86.2930.2001.Hansson,arXiv:0712.2174v1.Ouvry}). 
Several authors have also discussed the microscopic reason for the 
manifestation of FES \cite{PhysRevLett.80.1698.1998.Iguchi,PhysRevB.61.12757.2000.Iguchi,PhysRevLett.73.3331.1994.Murthy,PhysRevLett.74.3912.1995.Sen,PhysRevB.60.6517.1999.Murthy,NuclPhysB470.291.1996.Hansson,IntJModPhysA12.1895.1997.Isakov,PhysRevLett.86.2930.2001.Hansson,PhysRevLett.85.2781.2000.Iguchi,arXiv:0712.2174v1.Ouvry}. 

Iguchi and Sutherland \cite{PhysRevLett.85.2781.2000.Iguchi} showed that 
liquids of particles in three dimensions, interacting through long-range 
forces exchibit the nature of quantum liquids with FES, the characteristics 
of the FES being determined by the interaction. 

Murthy and Shankar \cite{PhysRevLett.73.3331.1994.Murthy} analysed a 
system of fermions in the Colagero-Sutherland model. The system has 
a constant density of states (DOS) (along the single particle energy axis) 
and has a total energy of 
\begin{equation}
E(\{n_i\}) = \sum_i \epsilon_i n_i + \frac{V}{2\sigma} N(N-1) , \label{EtotMS}
\end{equation}
where $n_i$ is the population of the single particle state of energy 
$\epsilon_i$, $\sigma=(\epsilon_i-\epsilon_{i-1})^{-1}$ (for any $i>0$) is 
the DOS, $V$ is the mean-field interaction potential, and $N$ is 
the total number of particles in the system. By redistributing in an 
uneven way the interaction energy between the particles of the system and 
associating to the level $i$ the quasi-particle energy 
\begin{equation}
\tilde\epsilon_i = \epsilon_i + V\sigma^{-1}\sum_{j=0}^{i-1} n_j ,
\label{Eredistribution}
\end{equation}
Murthy and Shankar obtained a gas with FES of parameter $\alpha=1+V$. 

A model which is similar to that of Murthy and Shankar 
\cite{PhysRevLett.73.3331.1994.Murthy} has been employed also in Refs. 
\cite{NuclPhysB470.291.1996.Hansson,IntJModPhysA12.1895.1997.Isakov,PhysRevLett.86.2930.2001.Hansson} 
to describe anyons on the lowest Landau level, coupled chiral particles on 
a circle, or interacting bosons in two-dimensions. 

In Refs. 
\cite{JPA35.7255.2002.Anghel,RomRepPhys59.235.2007.Anghel} I showed that the 
same model, with a slight generalization, can lead to a condensation, which is 
a first order phase transition. 

In this paper I will extend the method of Murthy and Shankar to systems 
of general DOS and any interaction potential, $V_{ij}$ (where $i$ and $j$ 
label the single particle states) and I will 
show that such systems lead to a more general manifestation of FES. 
While in the Murthy and Shankar model we have only \textit{direct} exclusion 
statistics (i.e. the exclusion statistics is manifested only in the subspace 
where the particles are inserted) of constant parameter, $\alpha$, here, in 
the general case, we shall have also \textit{mutual} statistics (acting from 
one subspace into another); therefore we shall have more complex 
parameters, denoted as $\alpha_{ij}$. 
I will calculate 
explicitely the parameters $\alpha_{ij}$ and I will prove that 
the mutual parameters ($\alpha_{ij}$, with $i\ne j$) are proportional 
to the dimension of the Hilbert subspace on which they act, 
verifying in this way the conjecture put forward in 
Ref. \cite{JPhysA.40.F1013.2007.Anghel}.

\section{FES in systems with interaction} \label{FESint}

\begin{figure}
\begin{center}
\resizebox{30mm}{!}{\includegraphics{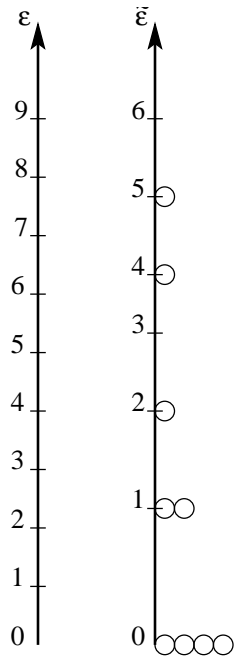}}
%\resizebox{30mm}{!}{\includegraphics{pics/MSbasic.eps}}
\end{center}
\caption{The single particle energy levels in the noninteracting 
system (left) and the corresponding quasiparticle enegy levels 
(right, $\tilde\epsilon_i=\epsilon_i + \sum_{j=0}^{i-1} V_{ij}n_j$) when 
there are four particles on level 0, two on level 1, and one particle on 
each of the levels 2, 4, and 5. In this particular case I chose 
$\epsilon_i=i$ and $V_{ij}=1/3$ for any $i,\ j$.}
\label{MSbasic}
\end{figure}

Let us generalize the model of Murthy and Shankar \cite{PhysRevLett.73.3331.1994.Murthy,NuclPhysB470.291.1996.Hansson,IntJModPhysA12.1895.1997.Isakov,PhysRevLett.86.2930.2001.Hansson}  by writing the total energy as 
\begin{equation}
E = \sum_i\epsilon_i n_i + \frac{1}{2}\sum_{ij}V_{ij}n_in_j 
\label{Etotgen}
\end{equation}
and the quasiparticle energies as 
\begin{equation}
\tilde\epsilon_i = \epsilon_i + \sum_{j=0}^{i-1} V_{ij}n_j +\frac{1}{2}V_{ii}n_i
\label{epstilgen}
\end{equation}
(see figure \ref{MSbasic}). 
To make the calculations and the physical implications as clear as 
possible, we assume that we have bosons in the systems--in this way we 
shall not have to worry about adding a unit to the direct exclusion 
statistics parameters. 
%An example of how $\tilde\epsilon_i$ is calculated is given in %figure \ref{MSbasic}, for $\epsilon_i=i/C$, $V_{ij}=1/3C$ ($i\ne j$), and $V_{ii}=0$. 
%
I will also assume that the system is large enough, so that the spectrum is 
(quasi)continuous, of the (generally not constant) DOS, $\sigma(\epsilon)$. 
Then, assuming that $V_{ij}$ depends only on the energies of the interacting 
particles, in Eq. (\ref{epstilgen}) I drop the subscript $i$ and I use 
$\epsilon$ as a variable, to write 
\begin{equation}
\tilde\epsilon = \epsilon + \int_0^\epsilon V(\epsilon,\epsilon') 
\sigma(\epsilon')n(\epsilon')\,d\epsilon'. 
\label{epstilgenint}
\end{equation}
In Eq. (\ref{epstilgenint}) I also ingnored the term 
$V(\epsilon,\epsilon)n(\epsilon)$. Although this term, for $\epsilon=0$, 
may cause a first order phase transition 
\cite{JPA35.7255.2002.Anghel,RomRepPhys59.235.2007.Anghel}, 
here I just want to emphasize the characteristics of the emerging FES and 
carring along this term in the calculations would be useless. 
I assume also that the function $\tilde\epsilon(\epsilon)$ is bijective, so 
that I can use freely its inverse, $\epsilon(\tilde\epsilon)$. Since 
$\tilde\epsilon(\epsilon)$ and $\epsilon(\tilde\epsilon)$ depend also 
on the populations of the energy levels below $\epsilon$ or below 
$\tilde\epsilon$, respectively, I shall use also the notations 
$\tilde\epsilon_{n(\epsilon'<\epsilon)}(\epsilon)$ and 
$\epsilon_{n(\tilde\epsilon'<\tilde\epsilon)}(\tilde\epsilon)$ whenever 
this will be needed for clarity. 

If I denote the density of states along the $\tilde\epsilon$ axis by 
$\tilde\sigma(\tilde\epsilon)$ and the number of particles between the 
energy levels $\tilde\epsilon_1$ and $\tilde\epsilon_2$, by 
$N(\tilde\epsilon_1,\tilde\epsilon_2)$, then we have the relation 
\[%\begin{equation}
N(\tilde\epsilon_1,\tilde\epsilon_2)\equiv\int_{\tilde\epsilon_1}^{\tilde\epsilon_2}
\tilde\sigma(\tilde\epsilon)n(\tilde\epsilon)\,d\tilde\epsilon
= \int_{\epsilon(\tilde\epsilon_1)}^{\epsilon(\tilde\epsilon_2)}
\sigma(\epsilon')n(\epsilon')\,d\epsilon',
\]%\end{equation}
where, obviously, $n(\tilde\epsilon)\equiv n[\epsilon(\tilde\epsilon)]$. 

To show the underlying FES character of the system, I use the coarse-graining 
of the energy axis $\tilde\epsilon$. 
I split the quasiparticle energy axis into 
intervals--$[\tilde\epsilon_0,\tilde\epsilon_1], 
\ldots,[\tilde\epsilon_{M-1},\tilde\epsilon_M], \ldots$--which are small, 
but still contain large enough numbers of particles and energy levels; 
the FES will be manifested between and within these intervals 
\cite{PhysRevLett.80.1698.1998.Iguchi,PhysRevB.61.12757.2000.Iguchi,PhysRevLett.73.3331.1994.Murthy,PhysRevLett.74.3912.1995.Sen,PhysRevB.60.6517.1999.Murthy,NuclPhysB470.291.1996.Hansson,IntJModPhysA12.1895.1997.Isakov,PhysRevLett.86.2930.2001.Hansson,PhysRevLett.85.2781.2000.Iguchi,arXiv:0712.2174v1.Ouvry}. 
To each $\tilde\epsilon_i$ it corresponds an 
$\epsilon_i\equiv\epsilon_{n(\epsilon<\epsilon_i)}(\tilde\epsilon_i)$. 
I rewrite Eq. (\ref{epstilgenint}) as a summation, 
\begin{equation}
\tilde\epsilon_M = \epsilon_M + \sum_{i=0}^{M-1} V(\epsilon_M,
\epsilon_{i})N(\tilde\epsilon_{i},\tilde\epsilon_{i+1}) ,
\label{tildeeps_sum_interv}
\end{equation}
where, based on the fact that the intervals 
$[\tilde\epsilon_i,\tilde\epsilon_{i+1}]$ are small and $V$ is assumed to be 
continuous in both variables, I used the approximation 
$V(\epsilon_M,\epsilon_i)\approx V(\epsilon_M,\epsilon_{i-1})$ for any 
$i<M$. Using this decomposition I calculate the FES parameters. 
\begin{figure}
\begin{center}
\resizebox{110mm}{!}{\includegraphics{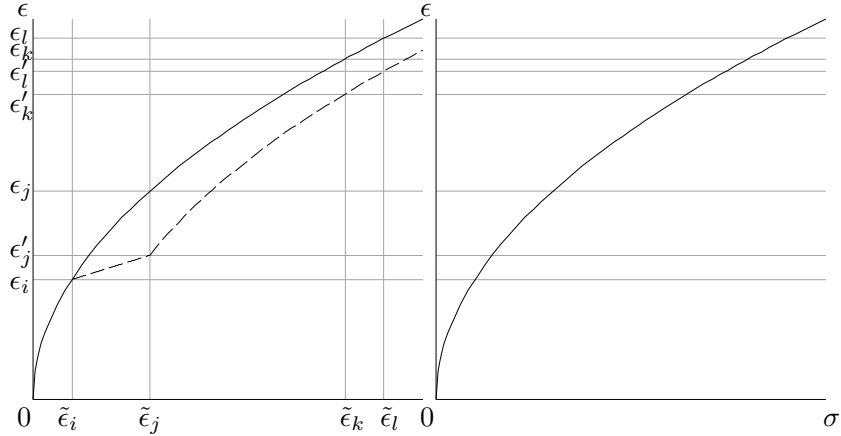}}
%\resizebox{110mm}{!}{\includegraphics{pics/insertion.eps}}
\end{center}
\caption{\textit{Left}: $\epsilon(\tilde\epsilon)$ before (solid curve) and 
after (dashed curve) the insertion of extra particles in the interval 
$(\tilde\epsilon_1,\tilde\epsilon_2)$. This changes the values 
of $\epsilon_1$, $\epsilon_2$, and $\epsilon_3$ into 
$\epsilon'_1$, $\epsilon'_2$, and $\epsilon'_3$, respectively. 
To the \textit{right} I draw the inverse of $\sigma(\epsilon)$, to 
emphasize the 
change of the number of states ($\int\sigma(\epsilon)\rmd\epsilon$) 
in each of the intervals after the insertion of particles.
Both plots are schematic and are used only to illustrate the principle of 
calculation.}
\label{insertion}
\end{figure}

First I calculate the \textit{direct} exclusion statistics parameter 
by adding $I_{M-1}$ particles in the interval 
$[\tilde\epsilon_{M-1},\tilde\epsilon_M]$. 
Since I hold fix $\tilde\epsilon_{M-1}$ and $\tilde\epsilon_M$, then 
$\epsilon_{M-1}$ and all the energy levels below it will also remain fix, 
while $\epsilon_M$ and all the energy levels above it will change. 
I calculate the change of single particle states in the interval 
$[\tilde\epsilon_{M-1},\tilde\epsilon_M]$ 
(see figure \ref{insertion}, with $i=M-1$ and $j=M$); I use the notation 
$\epsilon'_M\equiv\epsilon_{M,n(\tilde\epsilon'<\tilde\epsilon),I_{M-1}}(\tilde\epsilon)$, which is the value taken by $\epsilon_{M}$ after the insertion of the 
$I_{M-1}$ particles. The initial number of states in the interval 
$[\tilde\epsilon_{M-1},\tilde\epsilon_M]$ is 
$G(\tilde\epsilon_{M-1},\tilde\epsilon_M)=\int_{\epsilon_{M-1}}^{\epsilon_M}\sigma(\epsilon')\rmd\epsilon'$ and after the addition of particles it changes into 
$G'(\tilde\epsilon_{M-1},\tilde\epsilon_M)=\int_{\epsilon_{M-1}}^{\epsilon'_M}\sigma(\epsilon')\rmd\epsilon'$. 
So, to calculate the difference 
$\delta G(\tilde\epsilon_{M-1},\tilde\epsilon_M)=G'(\tilde\epsilon_{M-1},\tilde\epsilon_M)-G(\tilde\epsilon_{M-1},\tilde\epsilon_M)=\int_{\epsilon_{M}}^{\epsilon'_M}\sigma(\epsilon')\rmd\epsilon'$, 
I calculate first the change of $\epsilon'_M$: 
\begin{equation}
\tilde\epsilon_M = \epsilon'_M + V(\epsilon'_M,\epsilon_{M-1})I_{M-1} 
+ \sum_{i=0}^{M-1} V(\epsilon'_M,\epsilon_{i})
N(\tilde\epsilon_i,\tilde\epsilon_{i+1}) 
%\nonumber \\&=& \epsilon_M + \sum_{i=0}^{M-1} V(\epsilon_M,\epsilon_{i})N(\tilde\epsilon_i,\tilde\epsilon_{i+1})
\label{tildeeps_sum_interv_intro}
\end{equation}
If I denote $\delta\epsilon_M=\epsilon'_M-\epsilon_M$ and I expand 
$V(\epsilon_M,\epsilon_i)$ around $\epsilon_M$, I get from 
(\ref{tildeeps_sum_interv}) and (\ref{tildeeps_sum_interv_intro}) 
an equation for $\delta\epsilon$: 
\begin{equation}
\delta\epsilon_M =  \frac{- V(\epsilon_{M},\epsilon_{M-1})I_{M-1}}{1 + 
\frac{\partial V(\epsilon_{M},\epsilon_{M-1})}{\partial\epsilon_M}I_{M-1}
+ \sum_{i=0}^{M-1} \frac{\partial V(\epsilon_M,\epsilon_{i})}
{\partial\epsilon_M}N(\tilde\epsilon_i,\tilde\epsilon_{i+1})} 
\label{Eq.delta_eps}
\end{equation}
or, changing the summation into an integral, 
\begin{equation}
\delta\epsilon =  \frac{- V(\epsilon_M,\epsilon_{M-1})I_{M-1}}{1 + 
\frac{\partial V(\epsilon,\epsilon)}{\partial\epsilon_M}I_{M-1}
+ \int_{0}^{\epsilon(\tilde\epsilon)} \frac{\partial V(\epsilon,\epsilon')}
{\partial\epsilon}\sigma(\epsilon')n(\epsilon')\,d\epsilon'} .
\label{Eq.delta_eps_int}
\end{equation}
I look for linear effects, therefore I ignore the term proportional to 
$I_{M-1}$ from the denominator of equation (\ref{Eq.delta_eps_int}) and
I replace $V(\epsilon_M,\epsilon_{M-1})$ by $V(\epsilon_M,\epsilon_{M})$ 
(assuming that $V$ is continuous in both variables). 
Writing 
$\delta G(\tilde\epsilon_{M-1},\tilde\epsilon_M)=\delta\epsilon_M\cdot\sigma(\epsilon_M)\equiv\alpha_{\tilde\epsilon_M\tilde\epsilon_M}I_{M-1}$, I get 
%Therefore the insertion of $I_{M-1}$ particles in the 
%interval $[\tilde\epsilon_{M},\tilde\epsilon_{M-1}]$ changed the number 
%of states in the 
%interval by $\delta G_{M-1}=\sigma(\epsilon)\delta\epsilon$, giving a 
the \textit{direct} exclusion statistics parameter 
\begin{equation}
\alpha_{\tilde\epsilon\tilde\epsilon} =  \frac{V(\epsilon_M,\epsilon_M)
\sigma[\epsilon(\tilde\epsilon)]}{1
+ \int_0^\epsilon \frac{\partial V[\epsilon,\epsilon']}
{\partial\epsilon}\sigma(\epsilon')n(\epsilon')\,d\epsilon'} .
\label{alpha_eps_eps}
\end{equation}
Note that $\alpha_{\tilde\epsilon\tilde\epsilon}$ is identical to $\alpha$ 
calculated before \cite{PhysRevLett.73.3331.1994.Murthy,PhysRevLett.74.3912.1995.Sen,PhysRevB.60.6517.1999.Murthy,NuclPhysB470.291.1996.Hansson,IntJModPhysA12.1895.1997.Isakov,PhysRevLett.86.2930.2001.Hansson}  if 
$\partial V[\epsilon,\epsilon(\tilde\epsilon_{i})]/\partial\epsilon\equiv0$.

Now let's calculate the \textit{mutual} exclusion statistics parameters. 
For this I introduce $I_i$ particles in the interval 
$[\tilde\epsilon_{i},\tilde\epsilon_{i+1}]$ ($0\le i<M-1$). 
This will change all the energy levels 
$\epsilon_j$, of $j>i$ (see figure \ref{insertion}); let's denote the 
new values of $\epsilon_j$, $j>i$, by $\epsilon'_j$. 
Taking all these into account, I write 
%%
%\begin{equation}
%\tilde\epsilon = \epsilon' + V(\epsilon',\epsilon_i)I_i 
%+ \sum_{j=0}^{M-1} V[\epsilon',\epsilon(\tilde\epsilon_{j})]
%N(\tilde\epsilon_j,\tilde\epsilon_{j+1}) .
%\label{tildeeps_sum_interv_i}
%\end{equation}
%%
%To simplify the notations a bit, I will denote by $\epsilon_j$ the value 
%$\epsilon(\tilde\epsilon_j)$ in the 
%first configuration of particles (i.e. before introducing $I_i$) and 
%by $\epsilon'_j$ the value $\epsilon(\tilde\epsilon_j)$ in the configuration 
%with the $I_i$ particles included. Then Eq. (\ref{tildeeps_sum_interv_i}) 
%becomes 
%
\begin{eqnarray}
\tilde\epsilon &=& \epsilon' + V(\epsilon',\epsilon_i)I_i 
+ \sum_{j=0}^{i} V[\epsilon',\epsilon_{j}]
N(\tilde\epsilon_j,\tilde\epsilon_{j+1}) 
+ \sum_{j=i+1}^{M-1} V[\epsilon',\epsilon'_{j}]
N(\tilde\epsilon_j,\tilde\epsilon_{j+1}) .
\label{tildeeps_sum_interv_ip}
\end{eqnarray}
Expanding again $V(\epsilon,\epsilon')$ to the linear order in both variables, 
I get the equation for 
$\delta\epsilon_M^{(i)}\equiv(\epsilon'_M)^{(i)}-\epsilon_M$: 
\begin{eqnarray}
\delta\epsilon_M^{(i)}
%\left[1+I\frac{\partial V(\epsilon_M,\epsilon_i)}{\partial\epsilon_M} 
%+\sum_{j=0}^{M-1} \frac{\partial V(\epsilon_M,\epsilon_{j})}
%{\partial\epsilon_M}N(\tilde\epsilon_j,\tilde\epsilon_{j+1})\right] 
&=& -\frac{I_iV(\epsilon_M,\epsilon_i)+\sum_{j=i}^{M-1}\frac{\partial 
V(\epsilon_M,\epsilon_j)}{\partial\epsilon_j}
N(\tilde\epsilon_j,\tilde\epsilon_{j+1})\delta\epsilon_j}
{1+I\frac{\partial V(\epsilon_M,\epsilon_i)}{\partial\epsilon_M} 
+\sum_{j=0}^{M-1} \frac{\partial V(\epsilon_M,\epsilon_{j})}
{\partial\epsilon_M}N(\tilde\epsilon_j,\tilde\epsilon_{j+1})},
\label{Eq.delta_eps_i} 
\end{eqnarray}
where I used the superscript to indicate that the particles were 
inserted at $\tilde\epsilon_i$. 
The unknown quantities, 
$\delta\epsilon_j^{(i)}=(\epsilon_j')^{(i)}-\epsilon_j$, can 
be calculated recursively, 
starting from $j=i$, using first equation (\ref{Eq.delta_eps_int}) 
and then equation (\ref{Eq.delta_eps_i}). 
By doing so, we first notice that $\delta\epsilon_j^{(i)}$ 
\textit{is proportional to} 
$I_i$, for any $j$. Transforming both summations of equation 
(\ref{Eq.delta_eps_i}) into integrals and introducing the notation 
\begin{equation}
\fl
f(\tilde\epsilon_M,\tilde\epsilon_i) = \frac{\sum_{j=i}^{M-1}\frac{\partial 
V(\epsilon_M,\epsilon_j)}{\partial\epsilon_j}
N(\tilde\epsilon_j,\tilde\epsilon_{j+1})\delta\epsilon_j^{(i)}}{I_i} 
= \frac{\int_{\epsilon_i}^{\epsilon_M}\frac{\partial 
V(\epsilon_M,\epsilon')}{\partial\epsilon'}\sigma(\epsilon')n(\epsilon')
(\delta\epsilon')^{(i)}\rmd\epsilon'}{I_i} ,
\label{fepsepsp_def}
\end{equation}
I get the final equation for $\delta\epsilon$, 
\begin{equation}
\delta\epsilon(\tilde\epsilon_M,\tilde\epsilon_i) 
= -\frac{V(\epsilon_M,\epsilon_i)+ 
f(\tilde\epsilon_M,\tilde\epsilon_i)}
{1+\int_{0}^{\epsilon_M} \frac{\partial V(\epsilon_M,\epsilon')}
{\partial\epsilon_M}\sigma(\epsilon')n(\epsilon')\,d\epsilon'}I_i 
\label{int_eq_delta_eps}
\end{equation}
If we plug in equation (\ref{fepsepsp_def}) into equation 
(\ref{int_eq_delta_eps}), the later becomes an integral equation for 
$\delta\epsilon(\tilde\epsilon,\tilde\epsilon_i)$. 

Having now the expression for 
$\delta\epsilon_M^{(i)}$, we can calculate 
the change of the number of states in the interval 
$[\tilde\epsilon_{M-1},\tilde\epsilon_M]$: 
\begin{equation}
\delta G(\tilde\epsilon_{M-1},\tilde\epsilon_M) 
= \sigma(\epsilon_M)\delta\epsilon_M
-\sigma(\epsilon_{M-1})\delta\epsilon_{M-1} 
\approx \left.\frac{d\sigma(\epsilon)}{d\epsilon}\right|_{\epsilon_M} 
(\epsilon_M-\epsilon_{M-1})\delta\epsilon_M ,
\label{dim_var_i}
\end{equation}
where we ignored $\delta\epsilon_M-\delta\epsilon_M$, 
since $\delta\epsilon_M$ is itself a small quantity. 
Notice that because both, $\epsilon_{M-1}$ and $\epsilon_M$, vary 
at the insertion of particles at energies lower than $\epsilon_{M-1}$, 
the variation of the number of quasiparticle states in the 
interval $[\tilde\epsilon_{M-1},\tilde\epsilon_M]$ is proportional to 
$\epsilon_M-\epsilon_{M-1}$, i.e. is proportional to the dimension 
of the interval. 
%Because of this, the mutual exclusion statistics parameter, $\alpha_{\tilde\epsilon_M\tilde\epsilon_i}$ is proportional also to the dimension of the interval, as conjectured in Ref. \cite{JPhysA.40.F1013.2007.Anghel}. Indeed, 
Plugging equation 
(\ref{int_eq_delta_eps}) into equation (\ref{dim_var_i}) I obtain the 
\textit{mutual} exclusion statistics parameter, 
\begin{equation}
\alpha_{\tilde\epsilon_M\tilde\epsilon_i} = 
\frac{(\epsilon_M-\epsilon_{M-1})
\{V(\epsilon_M,\epsilon_i)+ 
f(\tilde\epsilon,\tilde\epsilon_i)\}}
{1+\int_{0}^{\epsilon(\tilde\epsilon)} \frac{\partial V(\epsilon,\epsilon')}
{\partial\epsilon}\sigma(\epsilon')n(\epsilon')\,d\epsilon'} 
\left[\frac{d\sigma(\epsilon)}{d\epsilon}\right]_{\epsilon(\tilde\epsilon)}
\end{equation}
One can see immediately that if $d\sigma(\epsilon)/d\epsilon=0$ for 
any $\epsilon$, as 
it was in the case of constant density spectrum, 
$\alpha_{\tilde\epsilon_M\tilde\epsilon_i}=0$ for any 
$\tilde\epsilon_M\ne\tilde\epsilon_i$ 
\cite{PhysRevLett.73.3331.1994.Murthy,PhysRevLett.74.3912.1995.Sen,PhysRevB.60.6517.1999.Murthy,NuclPhysB470.291.1996.Hansson,IntJModPhysA12.1895.1997.Isakov,PhysRevLett.86.2930.2001.Hansson}. 

Now we observe directly the surprising character of the mutual 
exclusion statistics, namely that it is proportional to 
the energy interval on which it acts, $(\epsilon_M-\epsilon_{M-1})$. 
In Ref. \cite{JPhysA.40.F1013.2007.Anghel} I showed that this characteristics 
is necessary to ensure the self-consistency of the FES formalism, 
especially in the thermodynamic limit. The method to calculate the 
particle population for such exclusion statistics parameters is 
also given there. 

\section{Conclusions}

Fractional exclusion statistics (FES) is usually considered as an 
``exotic'' type of statistics, manifested in special types of systems. 
Contrary to this belief in this paper, by analysing a system with a very 
general model of interaction between the constituent particles, I showed 
that FES is rather the rule than the exception. FES is manifested 
in general in interacting systems. Moreover, I calculated the FES parameters 
of the model gas and I showed that the \textit{mutual} 
exclusion statistics parameters are proportional to the subspace 
on which they act. This conclusion is also in contradiction with the usual 
definition of FES and therefore seems peculiar. But it is not so. 
In Ref. \cite{JPhysA.40.F1013.2007.Anghel} I showed that the typical 
definition of the mutual exclusion parameters leads to inconsistencies 
in the thermodynamics calculations and, in order to eliminate these 
inconsistencies, the exclusion parameters must have exactly the properties 
deduced here. 

\ack

I thank Dr. A. P\^arvan for motivating discussions and Dr. St. Ouvry for 
bringing to my attention Ref. \cite{arXiv:0712.2174v1.Ouvry}. Most part 
of this work 
was done at the Bogoliubov Laboratory of Theoretical Physics, 
JINR Dubna, Russia and I want to thank the stuff of the 
Laboratory, especially Dr. S. N. Ershov and Dr. A. P\^arvan, for 
hospitality. The work was partially supported by the NATO grant, 
EAP.RIG 982080. 

\section*{References}
%\bibliography{/home/dragos/general}
%\bibliographystyle{unsrt}

\end{document}